\documentclass[%
 aip,
 cha,%
 amsmath,amssymb,
preprint,
]{revtex4-1}

\usepackage{graphicx}
\usepackage{dcolumn}
\usepackage{bm}

\begin{document}


\title[Multistable Bursting]{Mechanism, Dynamics, and Biological Existence of Multistability in a Large Class of Bursting Neurons}

\author{Jonathan P Newman}
\email{jnewman6@gatech.edu}
\homepage{http://www.prism.gatech.edu/~jnewman6}
\author{Robert J Butera}
\email{rbutera@gatech.edu}
\altaffiliation[also at]{School of Electrical and Computer Engineering, Georgia Institute of Technology, Atlanta, Georgia, 30332-0250, USA}
\affiliation{Wallace H. Coulter Dept. of Biomedical Engineering, Georgia Institute of Technology, Atlanta, Georgia, 30332-0535, USA}

\date{\today}

\begin{abstract}
Multistability, the coexistence of multiple attractors in a dynamical system, is explored in bursting nerve cells. A modeling study is performed to show that a large class of bursting systems, as defined by a shared topology when represented as dynamical systems, are inherently suited to support multistablity. We derive the bifurcation structure and parametric trends leading to mulitstability in these systems. Evidence for the existence of multirhythmic behavior in neurons of the aquatic mollusc \textit{Aplysia Californica} that is consistent with our proposed mechanism is presented.  Although these experimental results are preliminary, they indicate that single neurons may be capable of dynamically storing information for longer time scales than typically attributed to non-synaptic mechanisms.
\end{abstract}


\maketitle

\begin{quotation}
Neurons that support bursting dynamics are a common feature of neural systems. Due to their prevalence, great effort has been devoted to understanding the mechanisms underlying bursting and information processing capabilities that bursting dynamics afford. In this paper, we provide a link between neuronal bursting and information storage. Namely, we show that the mechanism implicit to bursting in certain neurons may allow near instantaneous modifications of activity state that lasts indefinitely following sensory perturbation. Thus the intrinsic, extra-synaptic state of these neurons can serve as a memory of a sensory event.
\end{quotation}

\section{Introduction} \label{sec:intro}
Bursting is a dynamic state characterized by alternating periods of fast oscillatory behavior and quasi-steady-state activity. Nerve cells commonly exhibit autonomous or induced bursting by firing discrete groups of action potentials in time. Autonomously bursting neurons are found in a variety of neural systems, from the mammalian cortex \cite{chagnac1989} and brainstem \cite{butera1999I,butera1999II,purvis2007} to identified invertebrate neurons  \cite{plant1976,kandel1979}.

Multirhythmicity in a dynamical system is a specific type of multistability which describes the coexistence of two or more oscillatory attractors under a fixed parameter set. Multirhythmicity has been shown to occur in vertebrate motor neurons \cite{hounsgaard1988}, invertebrate interneurons \cite{lechner1996}, and in small networks of coupled invertebrate neurons \cite{kleinfeld1990}. Additionally, multirhythmicity has been demonstrated in models of intracellular calcium oscillations \cite{haberichter2001} and coupled genetic oscillators \cite{koseska2007}.

Multistable systems can act as switches in response to an external input. The
feasibility of multistability as an information storage and processing mechanism in neural systems has been widely discussed in terms of neural recurrent loops and delayed feedback mechanisms \cite{hopfield1982,foss1996,mackey1984}.  Theoretical studies have shown multirhythmic bursting behavior in a number of single neuron models \cite{canavier1993,butera1998} as well as in a model two-cell inhibitory (half-center oscillator) network \cite{matveev2007}. In biological neurons, it is possible these dynamics are employed as a short-term memory. In this report provide a general explanation for the existence of multirhythmic bursting in previous studies \cite{butera1998,butera2002,canavier1993} and the characteristics of a bursting neuron that allow multirhythmic dynamics. Additionally, we provide experimental evidence suggesting the existence of this behavior in several identified bursting neurons of the aquatic mullusc {\it Aplysia Californica}. 

\section{A simple parabolic bursting model} \label{sec:simpmodel}
Dynamical bursting systems are a subset of the singularly perturbed (SP) class of differential equations, 
\begin{eqnarray}
\dot{x} &=& f(x,y) , \label{eq:1a} \\
\dot{y} &=&  \epsilon g(x,y) , \, x \in \textbf{R}^m, \, y \in \textbf{R}^{n}, \label{eq:1b}
\end{eqnarray}
where $0 \leq \epsilon$ is a small parameter. Using singular perturbation methods \cite{rinzel1985,rinzel1987}, the dynamics of bursting models can be explored by decomposing the full system into fast and slow-subsystems:  Eq.~\ref{eq:1a} and  Eq.~\ref{eq:1b}, respectively. The slow-subsystem can act independently \cite{ermentrout1986}, be affected synaptically \cite{matveev2007}, or interact locally with the spiking fast-subsystem \cite{plant1976,canavier1993,butera1998,matveev2007} to produce alternating periods of spiking and silence in time. To examine the dynamical mechanism implicit to a bursting behavior, $y$ is treated as a bifurcation parameter of the fast-subsystem. This is formally correct when $\epsilon = 0$ and  Eq.~\ref{eq:1b} degenerates into an algebraic equation, but the assumption is reasonable when there is large time separation between fast and slow dynamics.

Using this technique, all autonomously bursting single neuron models displaying multirhythmic bursting in the literature \cite{canavier1993,butera1998} are topologically classified as the circle/circle type \cite{izikevich2000,izhikevich2007}; that is, their fast-subsystem is driven back and forth across a saddle node on invariant circle (SNIC) bifurcation to produce alternating spiking and silent states. These models can be reduced to a form that supports a topological normal SNIC to and from the active phase \cite{izhikevich2007}. This system is,
\begin{eqnarray}
\dot{v} &=& I + v^2 + u_1 , \label{eq:2a} \\
\dot{u_1} &=& -\alpha u_2 , \label{eq:2b} \\
\dot{u_2} &=& -\beta (u_2-u_1), \label{eq:2c}
\end{eqnarray}
where $I$ is a constant current, $\alpha$ and $\beta$ are small positive constants. Trajectories are reset after a voltage spike by $v = v_\mathrm{c}, v \leftarrow v_\mathrm{r}$ and $(u_1,u_2) \leftarrow (u_1+d_1,u_2+d_2)$ where $v_\mathrm{c} - v_\mathrm{r}$, $d_1$ and $d_2$ are discrete shifts in variables that account for the effects of a spike. Thus, the slow-subsystem defined by  Eq.~\ref{eq:2b} and  Eq.~\ref{eq:2c}, is a damped linear oscillator defining a stable focus when $\beta < 4\alpha$ or node when $\beta \geq 4\alpha$. Under the parameter sets used here, the slow-subsystem is a focus.  Eq.~\ref{eq:2a} is the fast-subsystem and is a quadratic integrate and fire neuron.

Eqs.~(\ref{eq:2a}, \ref{eq:2b}, \ref{eq:2c}) is a SP system for small $\alpha$ and $\beta$. When $u_1$ is used as a bifurcation parameter of the singular system,  Eq.~\ref{eq:2a}, a saddle node bifurcation occurs when $u_1 = u_\mathrm{sn} = -I$. When $u_1 > u_\mathrm{sn}$, $v \rightarrow \infty$ like $\tan(t)$ (see Appendix~\ref{appendix:mathmethods:QIF}). In the full system, when $u_1 < u_\mathrm{sn}$ trajectories slowly converge on the equilibrium point of the slow focus. When $u_1 > u_\mathrm{sn}$, trajectories of the slow subsystem are interrupted by spiking events when $v$ goes to $v_\mathrm{c}$ and $(u_1,u_2)$ are discretely modified. The existence of a bursting solution is reliant on the interaction between spiking events and a flow of the slow focus -- neither activity type can exist indefinitely when isolated. A necessary condition for the existence of a limit cycle that represents bursting or tonic spiking is that the equilibrium point of the slow-subsystem $(u_1^*,u_2^*)$ must satisfy $u_1^* > u_\mathrm{sn}$.

\begin{figure}
\includegraphics[width=\columnwidth]{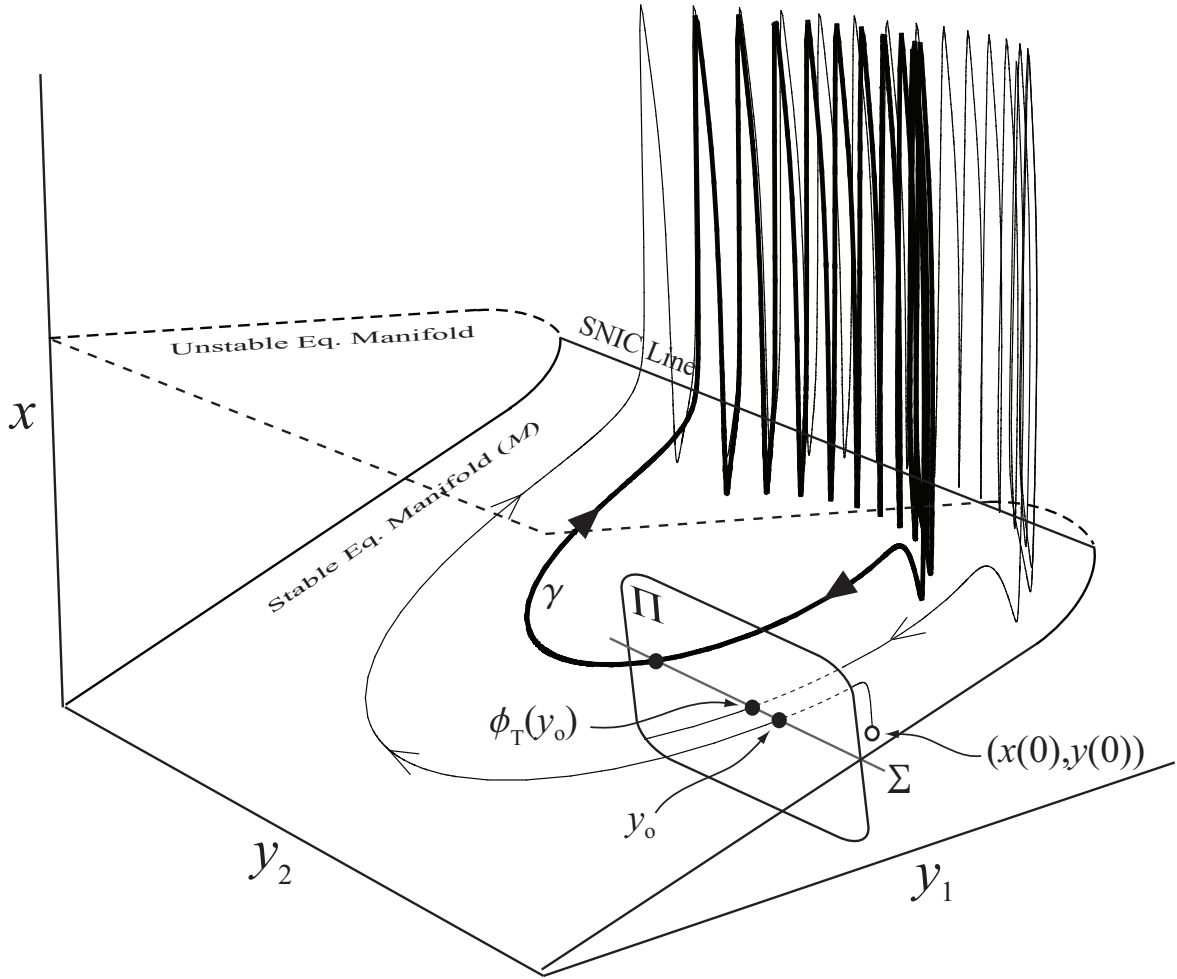}
\caption{A generic circle/circle bursting system projected into $\textbf{R}^{1+2}$. The dynamics of $x$ are much faster than those of $y$. At some points, trajectories are smashed onto the fast equilibrium manifold allowing return maps created from Poincar\'{e} sections of dimension $n-1$ to provide a complete dynamical description of the continuous system (see text for details).}
\label{fig1}
\end{figure}

A general aspect of SP systems of the form Eqs.~(\ref{eq:1a}, \ref{eq:1b}) is that if $M = \{ (x,y)|f(x,y)=0 \}$ is a stable equilibrium manifold of the fast subsystem on which $D_x f$ is non-singular, trajectories on $M$ follow the reduced field, $\dot{y} = g(h(y),y)$, where $h(y)$ is a function satisfying $f(h(y),y)=0$ \cite{guckenheimer1996}. With this in mind, consider a $m+n$ dimensional circle/circle bursting system (e.g. Eqs.~(\ref{eq:2a}, \ref{eq:2b}, \ref{eq:2c})). Let $\phi_t$ be a flow of the full system. On $M$, $h(y)$ is dependent on $y$ and static parameters. With $\epsilon = 0$, $\phi_t(x,y) = \phi_t(h(y),y)$ since the times-scales of the singular and slow-subsystem  are assumed to have infinite separation. Hence, if local cross section $\Pi \subset \textbf{R}^{m+n}$ of dimension $m+n-1$ is everywhere transverse to $\phi_t$, the condition $\phi_{\tau}(h(y_0),y_0)=(h(y_0),y_0)$ on $\Pi$ shows a $\tau$-periodic closed orbit. This condition reduces to $\phi_{\tau}(y_0)=y_0$ and fixed points of a $n-1$ dimensional map generated from a $n-1$ dimensional section on $M$ show closed orbits in $\textbf{R}^{m+n}$ (figure~\ref{fig1}). In the case of Eqs.~(\ref{eq:2a}, \ref{eq:2b}, \ref{eq:2c})), $n = 2$. Therefore a one-dimensional return map equivalent to the full system can be created using a one-dimensional section $\Sigma$.

\section{Bursting solutions result from counteracting dynamics}
\begin{figure}[ht]
\includegraphics[width=\columnwidth]{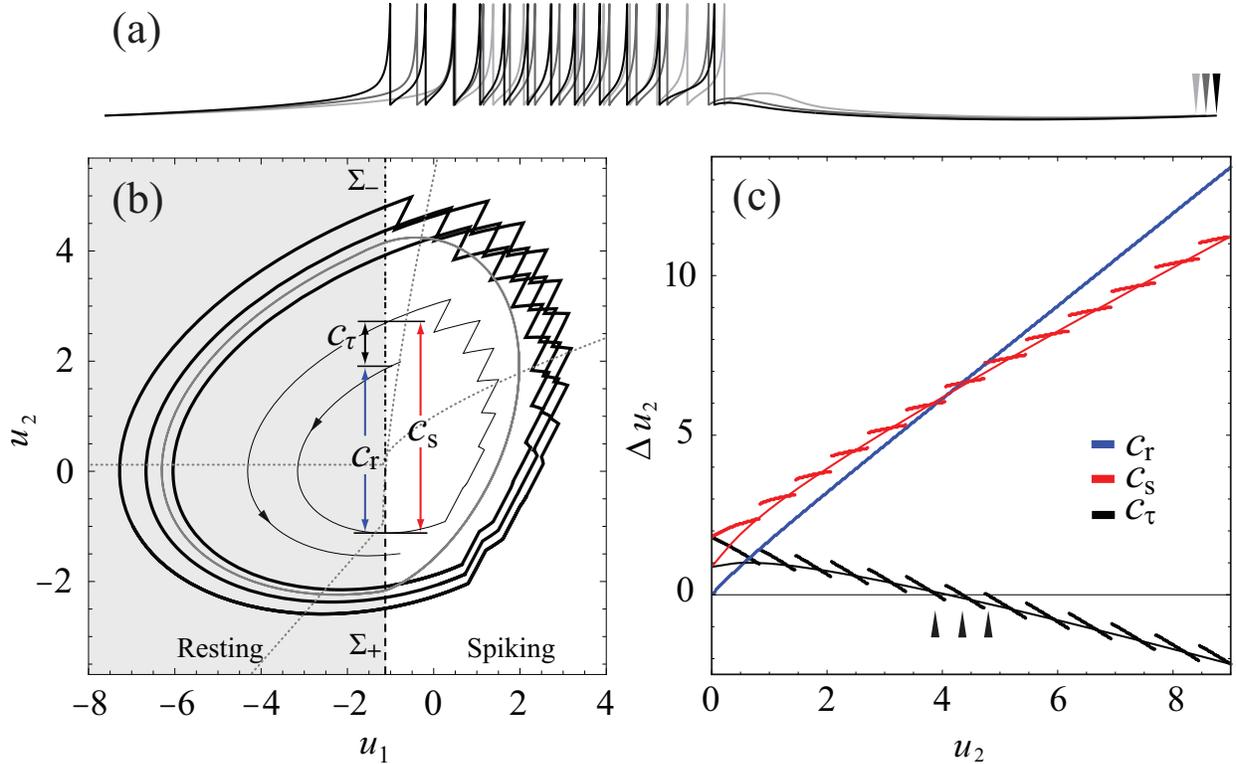}
\caption{Multirhythmic bursting produced by Eqs.~(\ref{eq:2a}, \ref{eq:2b}, \ref{eq:2c}). (a) Voltage traces showing a single period of three coexisting, limit cycles 10, 11, and 12 spikes. Note that the period of coexisting bursting solutions is different as indicated by three endpoint markers. (b) The corresponding limit cycles projected into the plane of the slow variables (thick) with the averaged solution shown in grey. Dotted grey lines are the nullclines of the averaged bursting system (see Eq.~\ref{eq:3a} and  Eq.~\ref{eq:3b}). Contraction metrics, $C_\mathrm{r}$, $C_\mathrm{s}$, and $C_{\tau}$ (Eqs.~(\ref{eq:contmet1}, \ref{eq:contmet2}, \ref{eq:contmet3}) are demonstrated on a portion of a trajectory of the full system spiraling outward (thin). The dashed line ($\Sigma_\mathrm{+}$) and dot-dashed line ($\Sigma_\mathrm{-}$) are the Poincar\'{e} sections used to define $G$ and $H$ (Eqs.~(\ref{eq:section_neg}, \ref{eq:section_pos}). (c) $C_\mathrm{r}$, $C_\mathrm{s}$, and $C_{\tau}$ for the multirhythmic system in (a). When the contraction of the slow focus is balanced by expansive spiking events, $C_{\tau} = 0$ and there is a fixed point of the map. Contraction metrics for the averaged system are thin solid lines. Parameters used here are $I=0.5$ ($I=1.2$ for averaged system), $\alpha=0.2$, $\beta=0.05$, $d_1=0.4$,$d_2=0.6$, $v_\mathrm{c}=10$, $v_\mathrm{r}=-1$.}
\label{fig2}
\end{figure}

Figure~\ref{fig2} indicates that discrete spiking events on a circle/circle bursting solution have a contraction-balancing action in phase space; they periodically force the stable focus, returning it to its initial condition after a single period, transforming the focal trajectory into a limit cycle. When trajectories containing \textit{different numbers} of spiking events return to their initial condition after a full revolution, multiple coexisting limit cycles are formed and multirhythmic bursting is achieved.

Consider multirhythmic bursting produced by Eqs.~(\ref{eq:2a}, \ref{eq:2b}, \ref{eq:2c}) in figure~\ref{fig2}. Let $M_\mathrm{c}$ represent the equilibrium manifold of  Eq.~\ref{eq:2a} parameterized by $u_1$. We recall from section~\ref{sec:simpmodel} that a $n-1 = 1$ dimensional section is needed to form a return map for the system defined by Eqs.~(\ref{eq:2a}, \ref{eq:2b}, \ref{eq:2c}) so long as that section is on $M_\mathrm{c}$. Therefore we can define this section as the line of saddle-nodes that divides the silent and spiking regimes of phase space, $\Sigma_\mathrm{c}= \{ (u_1,u_2) \in \textbf{R} \> | \> u_1 = u_\mathrm{sn} \}$. Consequently, a one dimensional recurrance map, $P :u_2 \rightarrow u_2$ is formed by intersections with the Poincar\'{e} section \footnote{Practically, the placement of $\Sigma_\mathrm{c}$ entails an error resulting from the assumption that fast and slow time scales are infinitely separated. In order to find the value of $u_1$ that truly separates spiking and resting states, we carried out numerous numerical simulations in the region of state space where contraction was balanced to determine the lowest value of $u_1$ that support a spiking event.},
\begin{eqnarray}
\Sigma_- \equiv \{ (u_1,u_2) \in \textbf{R} \> | \> u_1 \in \Sigma_\mathrm{c}, \dot{u_1} < 0 \}. \label{eq:section_neg}
\end{eqnarray}
To further explain how bursting and multirhythmicity arises in circle/circle systems, we divide the full return map $P$ into two components, $G$ and $H$ in terms of two Poincar\'{e} sections, $\Sigma_-$, and an additional section,  
\begin{eqnarray}
\Sigma_+ \equiv \{ (u_1,u_2) \in \textbf{R} \> | \> u_1 \in \Sigma_\mathrm{c}, \dot{u_1} > 0 \}.\label{eq:section_pos}
\end{eqnarray}
$G$ is a mapping from some intial condition, $u_2(0) \in \Sigma_-$ to a point on $\Sigma_+$. $H$ is then a mapping from  $G(u_2(0))$ back to  $\Sigma_-$ (figure~\ref{fig2}).  Thus, over the course of a periodic trajectory, $G$ accounts fo dynamics in the silent (contractive) region to the left of $\Sigma_\mathrm{c}$ and $H$ accounts for dynamics in the spiking (expansive) region to the right of $\Sigma_\mathrm{c}$. The functional composition of these maps,
\begin{eqnarray}
P(u_2) &=& H(G(u_2)),
\end{eqnarray}
is the discrete time Poincar\'{e} recurrence map as previously defined. Using $G$ and $H$, distance metrics,
\begin{eqnarray}
C_\mathrm{r}(u_2) &=& u_2 - G(u_2),\label{eq:contmet1} \\
C_\mathrm{s}(u_2) &=& G(u_2)-H(u_2),\label{eq:contmet2} \\
C_{\tau}(u_2) &=& C_s(u_2)-C_r(u_2),\label{eq:contmet3}
\end{eqnarray}
can be compared to describe contraction of trajectories in the resting, spiking, and combined regions of state space, respectively (see figure~\ref{fig2}). For points $u_2^{\ast} \in \Sigma_\mathrm{-}$  that $C_{\tau}(u_2^{\ast}) = 0$, the contraction of the resting portion of the trajectory is balanced by the net expansion of the spiking portion. Hence, $u_2^{\ast}$ are fixed points of $P$ and therefore show closed orbits in the full system.

Contraction and expansion of Eqs.~(\ref{eq:2a}, \ref{eq:2b}, \ref{eq:2c}) are clarified by examining the averaged slow-subsystem. This uses a near-identity change of variables to account for the time-averaged effect of fast spiking when $u_1>u_\mathrm{sn}$. Let $\varphi(t,u)$ be the limit cycle of the fast-subsystem defining $T$-periodic spiking. $\dot{u}=\epsilon g(\varphi(t,u),u)$ is the periodically forced slow-subsystem. By the Pontryagin-Rodygin theory \cite{pontryagin1960}, the averaged slow-subsystem for $u_1>u_\mathrm{sn}$ is defined as,
\begin{equation}
\dot{z}=\frac{\epsilon}{T(z)} \int_0^{T\!(z)}\!\! g(\varphi(t,z),z)dt, \label{eq:pont}
\end{equation}
where $z=u + \mathcal{O} (\epsilon)$. So, for our simple model, the averaged slow-subsystem is the switched system,
\begin{eqnarray}
\dot{z_1} &=& \left\{ \begin{array}{rcl}
-\alpha z_2 & \mbox{if}& z_1 \leq u_\mathrm{sn}\\
-\alpha z_2 + \frac{d_1}{T(z_1)} & \mbox{if}& z_1 > u_\mathrm{sn} \label{eq:3a}
\end{array}\right., \\
\dot{z_2} &=& \left\{  \begin{array}{rcl}
-\beta (z_2-z_1) & \mbox{if}& z_1 \leq u_\mathrm{sn}\\
-\beta (z_2-z_1) + \frac{d_2}{T(z_1)} & \mbox{if}& z_1 > u_\mathrm{sn} \label{eq:3b}
\end{array}\right.,
\end{eqnarray}
where,
\begin{widetext}
\begin{equation}
T(z_1)= \frac{1}{\sqrt{z_1+u_\mathrm{sn}}}(\arctan (\frac{v_\mathrm{c}}{\sqrt{z_1+u_\mathrm{sn}}})-\arctan (\frac{v_\mathrm{r}}{\sqrt{z_1+u_\mathrm{sn}}})), \label{eq:4}
\end{equation}
\end{widetext}
(see Figure~\ref{fig2} and Appendix~\ref{appendix:mathmethods}). Now opposing contraction dynamics are explicit between terms on the RHS of  Eq.~\ref{eq:3b} and  Eq.~\ref{eq:3b} when $z_1 > u_\mathrm{sn}$. 

\section{Conditions for multirhythmicity} \label{sec:conditions}
Non-monotonicity of $C_{\tau}$ is a necessary condition for multirhythmic bursting in circle/circle models because it means that $P$ may support several isolated contraction mappings. Fluctuations in the contraction of $P$ as measured by $C_{\tau}$ is the result of of near quantal spike addition in the active phase of bursting. Because the averaged system produces a monotonic $C_{\tau}$, it cannot generate a saddle-node bifurcation of closed orbits (SCO) and therefore cannot support multiple coexisting stable solutions. 

Consider a closed orbit of the simple model $\gamma$ projected into the plane $(u_1,u_2)$ containing $N$ spiking events. Dynamics on trajectories originating on the plane inside $\gamma$ is dominated by expansive spiking and they spiral outward. Dynamics on trajectories originating outside $\gamma$ is dominated by contraction of the focus and they spiral inward. However, when initial condition $(u_1(0),u_2(0))$ lies some critical distance outside $\gamma$, the arc length of the resulting trajectory past $u_\mathrm{sn}$ is long enough such that it contains $N+1$ spikes and its dynamics may become net expansive, causing it to spiral outward, indicating a `jump' in $C_{\tau}$. If this occurs, this trajectory forms the inner boundary of the trapping region of a second oscillatory attractor. Since in Eqs.~(\ref{eq:3a}, \ref{eq:3b}) quantal spiking is averaged over time, divisions between expansive and contractive annuluses about the equilibrium point are lost and $P$ is a single contraction mapping for all initial conditions  (figure~\ref{fig2}(c)).

Concentric basins of attraction in the plane must be divided by an unstable invariant set. In Eqs.~(\ref{eq:2a}, \ref{eq:2b}, \ref{eq:2c}), unstable periodic orbits are merely conceptual since spike addition occurs in a strictly discrete fashion. For a continuous system, unstable periodic orbits (UPO's) form separatixes for concentric basins of attraction. They are closed orbits that contain a dynamically unlikely attenuated action potential in the active phase. This can be seen in the biophysical models that support multirhythmic bursting \cite{canavier1993,butera1998}. UPO's in these systems correspond to the unstable fixed points of $P$ which separate contraction mappings.

\begin{figure}
\includegraphics[width=0.75\columnwidth]{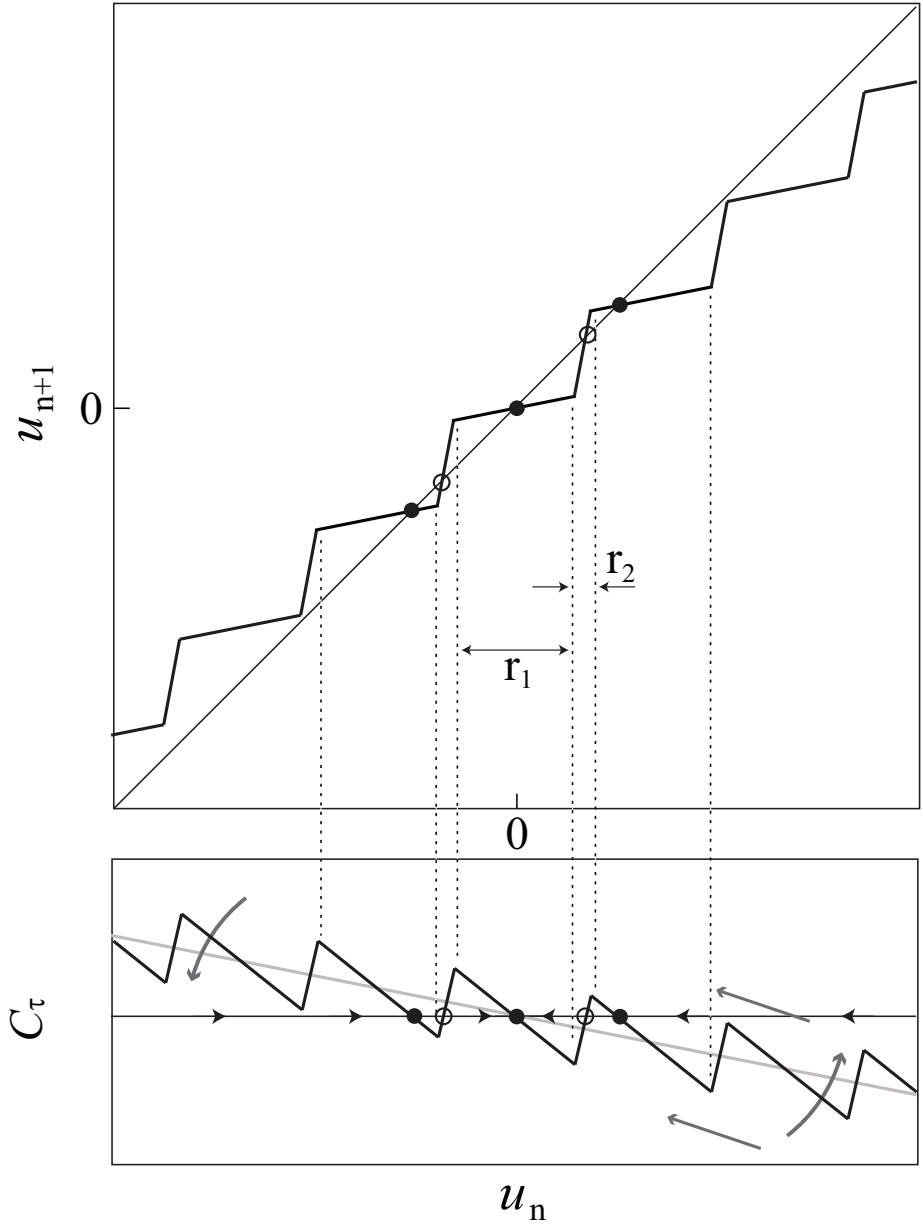}
\caption{An idealization of $P$ and $C_{\tau}$ for a circle/circle bursting model with a slow variable $u$. Stable and unstable fixed points are filled and open dots, respectively. As the slow subsystem becomes less dissipative and spiking events are capable of balancing contraction over larger areas of phase space. This effectively decreases the average slope of $C_{\tau}$ leading to the formation of multiple fixed points via SCO's (curved arrows).  Parametric changes that drive trajectories of the slow subsystem further past $u_\mathrm{sn}$ cause the number of spikes during an active phase to be increasingly sensitive to the initial condition. This effectively compresses $C_{\tau}$ increasing the number of separate contraction mappings via SCO's (straight arrows). $P$ has three separate contraction mappings over the three sets of preimages within the dotten lines.}
\label{fig3}
\end{figure}

$C_{\tau}$ is non-monotonic for Eqs.~(\ref{eq:2a}, \ref{eq:2b}, \ref{eq:2c}) so moving toward some parameter sets causes its local minima and maxima to cross zero, indicating the creation or annihilation of a stable/unstable orbit pair via a SCO. The damping ratio of the slow subsystem is $\zeta = \beta / 2\sqrt{ \alpha \beta}$. As damping is reduced in the slow subsystem by $\beta \rightarrow 0$ and $\alpha \rightarrow \infty$, dynamics move toward quasi-stable. This effectively reduces the average slope of $C_{\tau}$ pushing more of its extrema across zero causing SCO's to occur. Additionally, in circle/circle bursting systems, the firing rate past the SNIC scales as $\sqrt{b-b_\mathrm{sn}}$ where $b$ is the bifurcation parameter and $b_\mathrm{sn}$ is its value at the saddle node bifurcation (e.g  Eq.~\ref{eq:4} for the simple model). Since $|\dot{u_1}/\dot{u_2}|$ increases as $\alpha \rightarrow \infty$ and $\beta \rightarrow 0$, flow increases into $u_1$ following these parameter limits. This allows spike addition to occur more readily as a function of initial condition (figure~\ref{fig3}).

An interesting consequence of the mechanism underlying multirhythmic bursting in the simple model is that as $\alpha \rightarrow \infty$ and $\beta \rightarrow 0$ and when $d_1$ and $d_2$ are sizable, arbitrarily many stable bursting solutions can coexist.

\section{Classification of possible bursting behaviors} \label{sec:class}

From Section~\ref{sec:conditions} it is apparent that a circle/circle bursting system can be reduced to a one-dimensional map that switches between two modes. This can be verified for complicated models of circle/circle bursting which produce highly nonlinear return maps, but preserve an alternating, sawtooth-like structure \cite{butera1998}. 

To explore the dynamics possible under this constraint, we introduce a simple piecewise linear map acting on a slow variable $u$ that is depicted in figure~\ref{fig3}, top. Starting with a local contraction mapping symmetric about the origin, the map alternates in both directions between two moduli, $s_1$ and $s_2$, which act over sets of size $r_1$ and $r_2$, respectively. Since the map defined by $u_{n+1} = s_1u_n$ is a contraction mapping, we confine $|s_2| > 1$ so that it is possible to have multiple basins of attraction. The average slope of the map is given by $s_\mathrm{ave} = \frac{s_1 r_1 + s_2 r_2}{r_1 + r_2}$. $|s_{ave}| < 1$ requires the existence of at least one attractor under the map. Compared to $P$, the map is translated so that a central fixed point is located at the origin. Therefore, we consider the fixed point at the origin a nominal bursting solution with $N$ spikes. Jumps to different contraction mappings indicate jumps to new bursting solutions with $N\pm k, \, k = 1,2,3,\dots$ spikes in the active phase.

\begin{figure}
\includegraphics[width = \columnwidth]{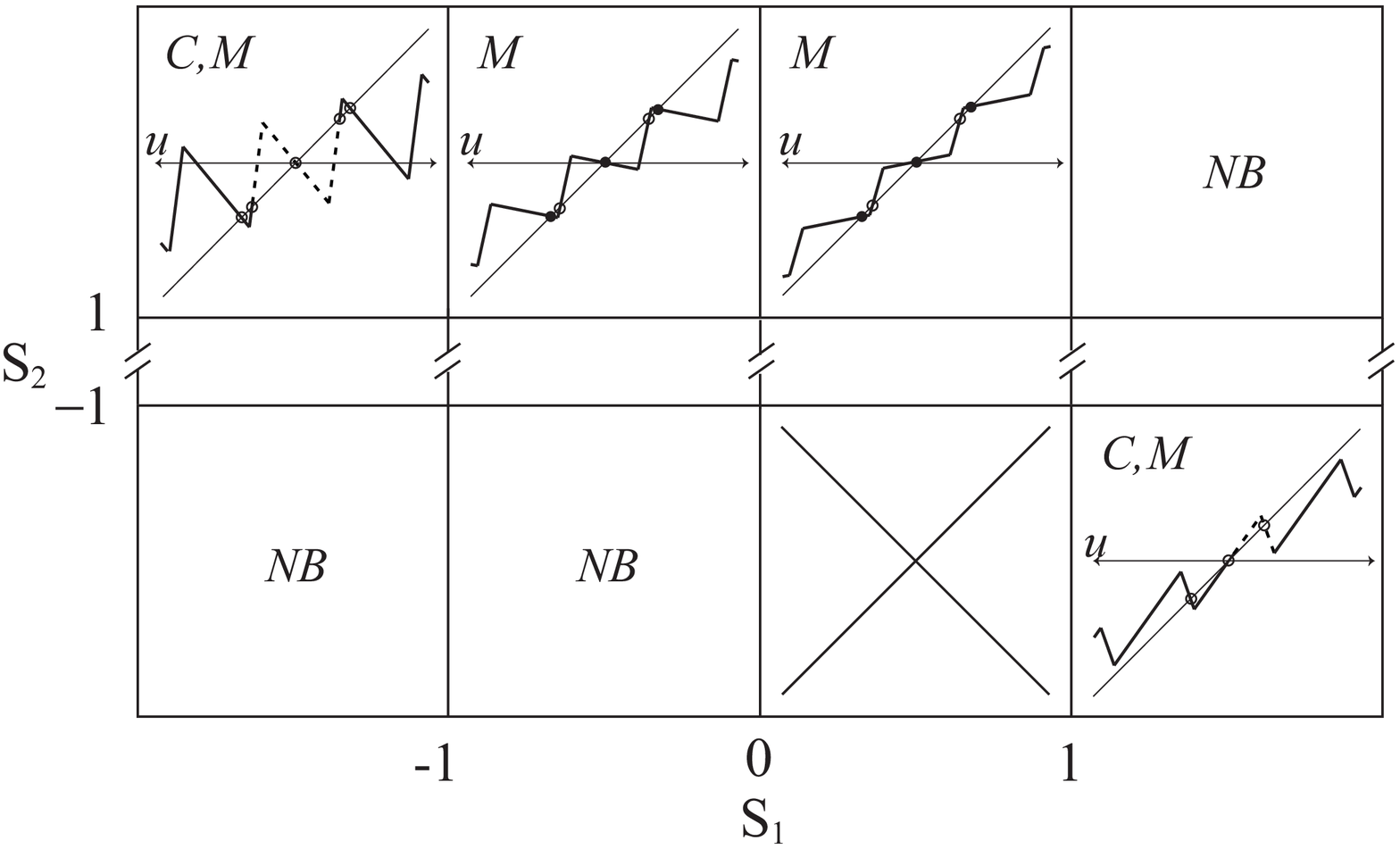}
\caption{Examples of the maps corresponding to the different parameter sets outlined in table~\ref{tab1}. Qualitative dynamics are classified by the following symbols: \textit{M} -- may support multirythmicity, \textit{C} -- chaotic, and \textit{NB} -- non-biological. Dashed sections of the maps show a single link in a chain of chaotic maps (sawtooth or tent map). The block for $0 < s_1 < 1, \, s_2 < -1$ is crossed out because the resulting map is not well defined.}
\label{fig4}
\end{figure}

Now the limiting behavior of circle/circle bursting can be explored by inspecting the map for different values of $s_1$ and $s_2$. This analysis is carried between figure~\ref{fig4} and table~\ref{tab1}. We note that the range of behaviors described by this simple system account for all the dynamics reported in previous multirhythmic bursting studies \cite{butera1998,canavier1993} including the coexistence of chaotic attractors. However, those models are not topologically conjugate to our piecewise linear map since they can produce combinations of limit-sets seen in different parameter regimes of the one dimensional system under a single parameter set (e.g. the coexistence of a limit cycle and a strange attractor).

\begin{table}
\caption{\label{tab1}The qualitative behavior of the piecewise linear map shown in figure~\ref{fig4} is described for each parameter regime. The top section describes why some parameter values are irrelevant to our analysis. The bottom section describes relevant parameter ranges that can fully account for the dynamics witnessed in prior multirhythmic bursting models.}
\begin{tabular*}{\columnwidth }{l | l| p{5 in}}
\hline\hline
$s_1$&$s_2$&Dynamics\\
\hline
-- & $|s_2| < 1 $ & $s_2$ must have modulus greater than 1\\
$|s_1| < 1 $ &  $s_2 < -1$ & Non-biological because $s_{ave} < 0$\\
$s_1 > 1 $ &  $s_2 > 1$ & Non-biological because $s_{ave} > 1$ indicating that the system is a repeller\\
$0 < s_1 < 1 $ &  $s_2 < -1$ & Map is not well defined\\
\hline
$s_1 < -1 $ & $s_2 > 1$ & Chaotic and possibly multirhythmic since the full map is formed by a chain of sawtooth maps with only unstable fixed points \\
$-1 < s_1 < 0 $ & $s_2 > 1$ & Attractors are oscillatory fixed points and system is possibly multirhythmic\\
$0 < s_1 < 1 $ & $s_2 > 1$ & Attractors are non-oscillatory fixed points and system is possibly multirhythmic\\
$s_1 > 1 $ & $s_2 < -1$ & Chaotic and possibly multirhythmic since the full map is formed by a chain of tent maps with only unstable fixed points \\
\hline\hline
\end{tabular*}
\end{table}

\section{Multirhythmic bursting in biological neurons} \label{sec:bio}

\begin{figure}
\includegraphics[width = \columnwidth]{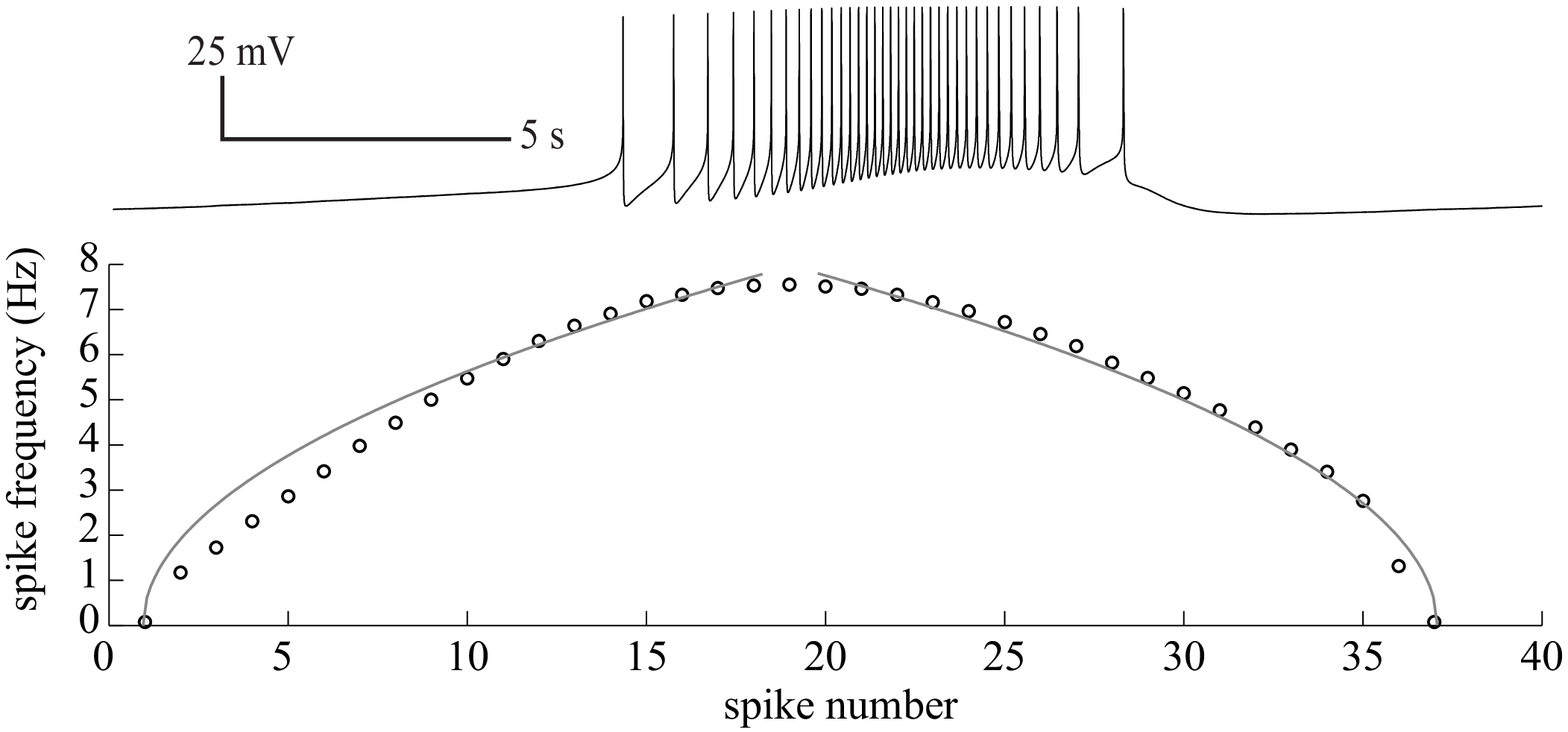}
\caption{Early studies of R15 deemed the neuron a `parabolic' burster since the spike-period profile resembles a parabola. However, it is now known that root scaling (grey line) better describes the frequency profile of this shape since after a SNIC the period of the resulting limit cycle scales as $1 / \sqrt{\lambda}$ (equation \ref{eq:4} for the simple model). This figure confirms this behavior in an \textit{in-situ} recording of R15. The instantaneous spike frequency during the active phase is shown with open circles.}
\label{fig5}
\end{figure}

We now provide evidence for the existence of multirhythmicity and multirhythmic bursting in invertebrate interneurons and neurosectretory cells. Cells R15, L3-L6, and L10 are identified neurons (neurons that are preserved animal to animal) located in the abdominal ganglion of \textit{Aplysia Californica}. These neurons burst spontaneously in-situ with or without synaptic isolation. Additionally, they display the `parabolic' bursting type as distinguished by two key features. First, they are characterized by root scaling in firing rate during the active phase (figure~\ref{fig5}). Secondly, action potentials during bursting have after-hyperpolarizations that are lower than the voltage threshold for the active phase, ruling out a hysteresis mechanism for bursting \cite{rinzel1985}. Therefore we can conclude the circle/circle type mechanism for bursting well describes their dynamics. 

R15 is the most heavily studied of the aforementioned neurons due to its particularly stable activity and long burst period \cite{kandel1979}. Numerous models \cite{plant1976,rinzel1987,canavier1991,bertram1993,butera1995} have attempted to describe the mechanisms underlying bursting in this cell and many of these models are of the circle/circle type \cite{izikevich2000}. One class of models \cite{butera1995,butera1998,canavier1991,adams1985,rinzel1987} postulates that the slow variables underlying bursting are:  are (1) the degree of voltage-gated activation of a slow, inward Ca$^{2+}$ current, I$_{SI}$ and (2) intracellular calcium concentration. Activation of I$_{SI}$ increases intracellular Ca$^{2+}$ and I$_{SI}$ displays voltage dependent activation and Ca$^{2+}$ dependent inactivation. 

We used a traditional `current-clamp' technique to control applied current across the cell membrane while monitoring the membrane potential. Since I$_{SI}$ and intracellular Ca$^{2+}$ concentration are directly and indirectly voltage dependent, current perturbations that change the membrane potential may influence these variables such that a multirhythmic cell is forced to switch attractors. Specifically, since our hypothesized mechanism of attractor switching involves spike addition, we aimed to change the number of action potentials in the active phase of bursting. 

\begin{figure*}
\includegraphics[width= \columnwidth]{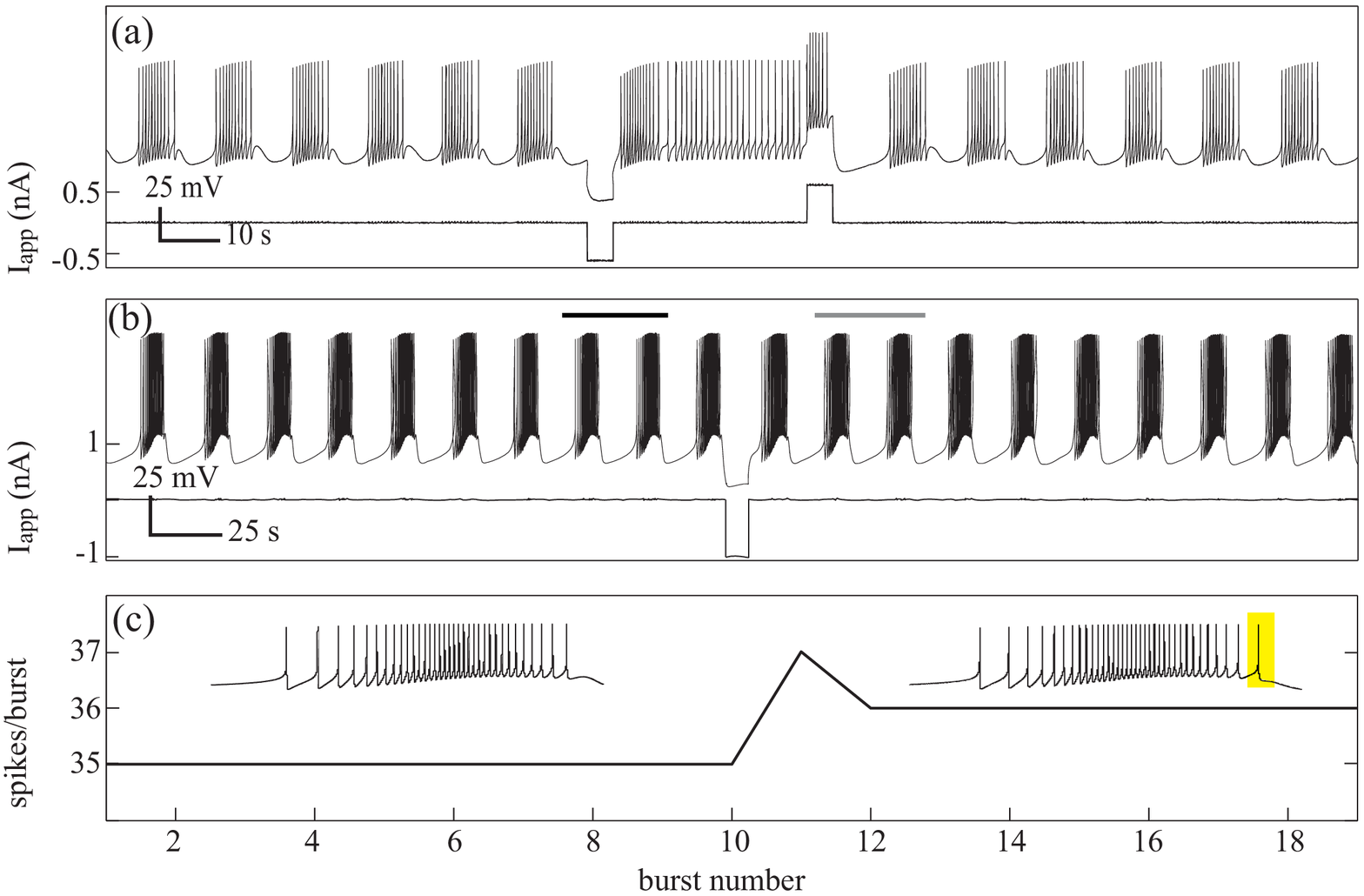}
\caption{Multirhythmicity in bursting neurons of \textit{Aplysia's} abdominal ganglion. (a) A `left-upper-quadrant' cell, most likely L3, showing bistability between bursting and tonic spiking. This behavior is predicted in biophysical models \cite{canavier1993,canavier1994} and was shown previously in R15 \cite{lechner1996}. Fundamentally, it is explained by the one-dimensional map $P$ (section~\ref{sec:class}) with $N=11$ spikes per burst in the nominal (non-perturbed) mode and $N=1$ spike per burst after the current pulse. Applied current pulses likely manipulate intracellular Ca$^{2+}$ concentrations, effectively moving the trajectory about in the plane of the slow variables, and allowing an attractor switch. (b) R15 showing bistability between two bursting solutions. Although the current pulse appears to have little effect, (c) shows the amount of spikes in the active phase of each burst. After the perturbation is applied, the number of spikes per burst increases by two, and then relaxes onto the new attractor containing 36 spikes per burst. The active phase for pre- and post-perturbation is shown with the added spike highlighted.}
\label{fig6}
\end{figure*}

Models of R15 predict that outward current perturbations hyperpolarize the cell from its spiking threshold and when the cell is released from this hyperpolarized state it spikes vigorously (`rebounds'), leading to spike addition. Since activation of I$_{SI}$ is depolarization dependent, prolonged hyperpolarization completely deactivates Ca$^{2+}$ channels supporting I$_{SI}$. Additionally, in the absence of an inward Ca$^{2+}$ flux, the intracellular Ca$^{2+}$ concentration decreases due to cytosolic buffering \cite{canavier1991} and the ion-channels supporting I$_{SI}$ are de-inactivated. When the inhibitory current is released the resulting depolarization quickly activates $I_{SI}$ allowing it to pass inward Ca$^{2+}$ current which maintains a high membrane potential causing spikes. Since intracellular Ca$^{2+}$ was initially very low, I$_{SI}$ must pass more Ca$^{2+}$ than usual before it is inactivated, resulting in a prolonged active phase. A depolarizing current has an opposing effect. Using 2 to 8 second, -0.5 to -2.5 nA hyperpolarizing current perturbations during the silent phase of bursting to add spikes to the subsequent active phase, we were able to show evidence for a multirhythmic behavior in four of eight bursting cells (see Appendix~\ref{appendix:expmethods} for details of the experimental method). Figure~\ref{fig6} shows recordings in two different bursting neurons from separate animals that display a multirhythmic behavior along with the current perturbations used to induce an attractor switch.

\begin{figure}
\includegraphics[width=\columnwidth]{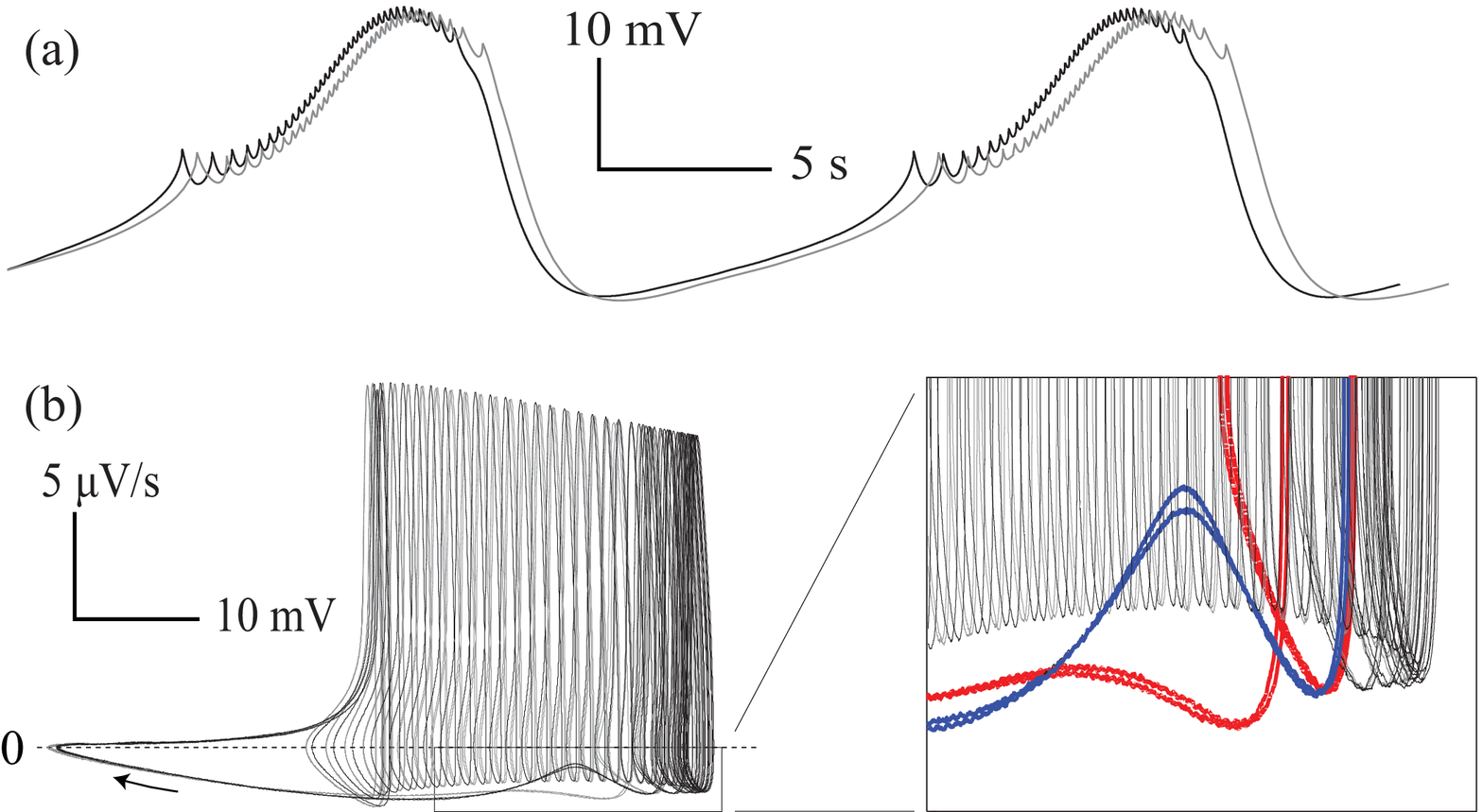}
\caption{Attractor reconstruction from experimentally derived voltage traces. (a) Two bursts preceding and subsequent to stimulation were used for a comparative attractor reconstruction (bars in figure~\ref{fig6}(b) denote the bursts used). A single burst period was defined by a the time between rising edges through -70 mV in the smoothed voltage trace. Voltage time series were low pass filtered using an exponentially weighted moving average with a width of 2 seconds to prevent spiking events from dominating the reconstructed attractor. Note the constant phase advance of one trace compared to the other, which is predicted by our model in figure~\ref{fig2}. (b) Voltage plotted against its estimated derivative. This attractor shape can be directly compared to biophysical models of neurons which rely on a circle/circle bursting mechanism. A zoomed portion of the reconstruction is shown to highlight the existence of a separatix dividing the two attractors, an unstable closed orbit that contains an attenuated spike. The final portion of the active phase is thickened for each attractor to make this clear.}
\label{fig7}
\end{figure}

Section~\ref{sec:conditions} predicts that an attractor switch in a multirhythmic burster will be characterized by spike addition or deletion and an increase or decrease in burst period, respectively. \ref{fig7} presents a comparison of slow-wave activity in \ref{fig6}(b) for two bursts prior and subsequent to the stimulus. Because coexisting bursting solutions are concentric in our models, the path-length of the solutions must differ resulting in distinct periods for each solution (see figure~\ref{fig2}(a)). Note the existence of a constant phase advance developed after spike addition shown in figure~\ref{fig7}(a) satisfies this prediction. In an attempt to elucidate the shape of the two coexisting attractors we plotted voltage versus its estimated derivative in figure~\ref{fig7}(b). Our multirhythmic model predicts that an unstable bursting solution containing an attenuated spike separates stable solutions. This separatix appears to exist, separating one attractor from the other it proceeds for an additional spike in figure~\ref{fig7}(c).

One may argue that the current perturbations we provide are too large both temporally an in terms of current amplitude to resemble a synaptic effect. However, the volume of charge moved across the membrane is a very superficial definition of synaptic efficacy. Smaller synaptic currents that rely on specific charge carriers (e.g. synaptic activation of calcium-specific channels) can be more effective in terms of altering specific neuron dynamics than our large chloride based currents. In the future, it would be interesting to repeat our experiment with direct perturbations to the hypothesized slow-subsystem via intracellular calcium uncaging. 

\section{Possible implications of multirhythmic bursting} \label{sec:implications}

Since the time scales associated with short term memory, for example the multirhythmic motor memory suggested in \cite{hounsgaard1988}, are shorter than those associated with morphological synaptic plasticity, it is possible that neural systems employ of some activity dependent multistability as a memory. Most proposed mechanisms of dynamic multistability in neural systems rely on some delayed feedback mechanism in a small neural circuit as the driving force in the creation of multiple coexisting attractors \cite{foss1996,mackey1984,hopfield1982}. Here we have shown that a sufficiently underdamped slow subsystem in a single circle/circle bursting neuron is enough to ensure the existence of multirhythmic behavior.

\begin{figure}
\includegraphics[width=\columnwidth]{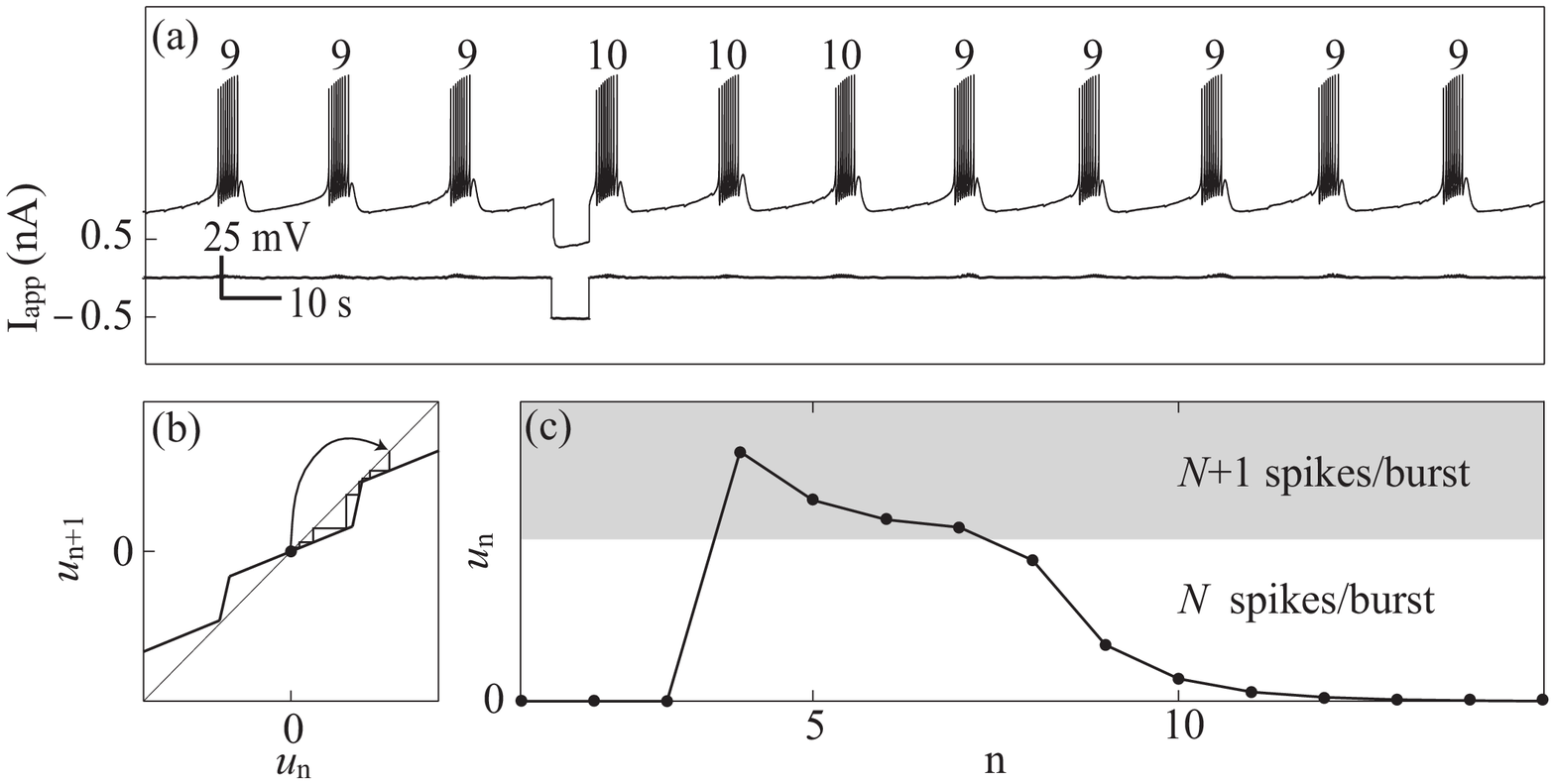}
\caption{`Fading' multirhythmicity in L4 of \textit{Aplysia's} abdominal ganglion. (a) Voltage trace and corresponding current perturbation. The three bursts following the perturbation contain an extra spike. This results from forcing the trajectory to pass through a bottleneck where dynamics are close to fixed which causes a lasting effect on bursting without the system actually being multirhythmic. (b) Fading multirhythmicity using the piecewise linear representation of $P$. Note that the map has areas that are close to forming fixed points. A trajectory can temporarily become caught here allowing a temporal amplification of a small perturbation (curved arrow). (c) Time series of the trajectory in (c). The ruins of an SCO allows the system to maintain $N+1$ spikes per burst for a time before falling back into the true attractor with $N$ spikes per burst.}
\label{fig8}
\end{figure}

We chose to show the existence of this phenomenon using invertebrate neurons because of their relative ease of experimental manipulation. However, as is put forth in sections \ref{sec:conditions} and \ref{sec:class}, the dynamical mechanism underlying multirhythmic bursting is general and the multirhythmic regime occupies a non-finite range of parameter space; circle/circle bursting systems appear inherently suited for short term information storage. Circuits have been proposed that take advantage of this fact \cite{butera2002}.  Even if a biological bursting neuron is not truly multistable, the shape of $P$ indicates that the system will always have a non-monotoncity in the contraction of the vector field about the attractor (in $C_{\tau}$). This allows temporal amplification of perturbations to the system, even if it has only one true attractor (figure~\ref{fig8}). 

In this report, we have proposed a basic dynamical mechanism for the existence of multirhythmic bursting in the biophysical models that previously demonstrated this behavior. We then explored the range of dynamics possible in circle/circle type bursting systems. Finally, we provided evidence for the existence of multirhythmic bursting in biological neurons. These experiments demonstrate a response to perturbations that is consistent with the dynamical model described herein. Further work concerning the nature of multirhythmic bursting neurons embedded within neural circuits is necessary to understand the importance of these findings in terms of short term memory.

\section{Acknowledgments} \label{sec:acknowledge}
The authors thank Astrid Prinz, Jianxia Cui, and Hiroki Sayama for helpful comments.  J. Newman was supported by Georgia Tech's NSF IGERT program on Hybrid Neural Microsystems (DGE-0333411) and an NSF Graduate Research Fellowship. Experimental and modeling work was also supported by grants from the NSF (CBET-0348338) and NIH (R01-HL088886), respectively, to R. Butera.

\appendix
\section{Mathematical details of the simple circle/circle bursting model} 
\label{appendix:mathmethods}
\subsection{The quadratic integrate and fire neuron model}
\label{appendix:mathmethods:QIF}

Consider the quadratic integrate and fire model, Eq.~\ref{eq:2a}. When $u_1$ is treated as a parameter, 
\begin{equation}
\dot{v} = b_{\mathrm{sn}}+v^2, \label{eq:Append1}
\end{equation}
where $b_{\mathrm{sn}} = I + u_1$. A saddle node bifurcation occurs when $\sqrt{b_{\mathrm{sn}}} = 0$ indicating that the roots of the RHS of  \ref{eq:Append1} have coalesced. When $b_{\mathrm{sn}} < 0$, these roots are real, $v_{\mathrm{eq}} = \pm \sqrt{b_{\mathrm{sn}}}$, and are the equilibria of  of  \ref{eq:Append1} . When $b_{\mathrm{sn}} > 0$ 
\begin{equation}
    v(t) = \sqrt{b_{\mathrm{sn}}}\tan(\sqrt{b_{\mathrm{sn}}}(t+t_0)), \label{eq:Append2}
\end{equation}
which is an exact solution for the membrane potential $v(t)$.

From this we can calculate the periodicity of spiking once the neuron has moved to the tonic regime by finding the time for $v(t)$ to go from $v_r$ to $v_c$. Let $t_1+t_0$ be the instant that a spike begins such that $v(t_1) = v_r$ and $t_2+t_0$ be the instant that a spike terminates such that $v(t_2) = v_c$. From \ref{eq:Append2},
\begin{eqnarray}
v_r &=& \sqrt{b_{\mathrm{sn}}}\tan(\sqrt{b_{\mathrm{sn}}}(t_1+t_0)) \label{eq:Append3}\\ 
v_c &=& \sqrt{b_{\mathrm{sn}}}\tan(\sqrt{b_{\mathrm{sn}}}(t_2+t_0)).
\label{eq:Append4}
\end{eqnarray}
Solving for $t_1$ and $t_2$,
\begin{eqnarray}
t_1 &=& (\arctan(\frac{v_r}{\sqrt{b_{\mathrm{sn}}}})/\sqrt{b_{\mathrm{sn}}}) - t_0,
\label{eq:Append5}\\
t_2 &=& (\arctan(\frac{v_c}{\sqrt{b_{\mathrm{sn}}}})/\sqrt{b_{\mathrm{sn}}}) - t_0.
\label{eq:Append6}
\end{eqnarray}
Therefore, the inter-spike period (the period of a fast-oscillation) is given by $T = t_1 - t_2$ which matches equation Eq.~\ref{eq:4}.

\subsection{Average bursting dynamics}
\label{appendix:mathmethods:Average}

The slow subsystem of the simple model, Eqs.~(\ref{eq:2b}, \ref{eq:2c}), is  $T$-periodically perturbed. Each time an action potential occurs in the fast subsystem,  Eq.~\ref{eq:2a}, there is an instantaneous modification of the slow variables. Let $d_i$ represent the magnitude of this modification on a single slow variable, $u_i$, for each spike. Finding the average contribution of these modifications on the time derivative of $u_i$ follows from the insertion of a delta function into Eq.~\ref{eq:pont},
\begin{equation}
\frac{1}{T(u_1)} \int_0^{T(u_1)} d_i \delta(t-T(u_1)) dt = \frac{d_i}{T(u_1)}\label{eq:Append9}.
\end{equation}
The time constant, $\epsilon$, is dropped in this calculation because discrete changes to the slow variables are completely removed from contributions of time constants in the continuous dynamics defined by  \ref{eq:2b} and \ref{eq:2c}. The validity of this result is confirmed by noting that,
\begin{equation}
\int_{0}^{T(u_1)} \frac{d_i}{T(u_1)} dt = \frac{d}{T(u_1)}T(u_1) = d_i. \label{eq:Append8}
\end{equation}
Therefore, when $d_i/T(u_1)$ is integrated over a single period, it produces an equivalent continuous change in the slow variables as the discrete event does instantly per single period.

\section{Experimental materials and methods} \label{appendix:expmethods}

\textit{Aplysia californica} were purchased from University of Miami Aplysia Resource Facility (Miami, FL) and maintained in a tank with filtered artificial sea water (ASW; Instant Ocean, Burlington, NC) at a temperature of about 19$^\circ$C until used. Each animal was anesthetized prior to experimentation via injection about 50\% of the animal's weight of isotonic MgCl$_2$: 71.2 g MgCl$_2$ in 1 L of ASW solution containing (in mM) 460 NaCl, 10 KCl, 11 CaCl$_2$, 30 MgCl$_2$, 25 MgSO$_4$, and 10 HEPES (4-(2-hydroxyethyl)-1-piperazineethanesulfonic acid, pH 7.6). The animal was then pinned to a large dissection dish and opened rostrally to caudally along the dorsal midline. The abdominal ganglion was excised and pinned dorsal side up in a dish coated with Sylgard (Dow Corning, Midland, MI) in a saline solution of 30\% isotonic MgCl$_2$ and 70\% ASW. The sheath of connective tissue covering the neurons of the ganglion was removed with fine scissors and fine forceps. The desheathed abdominal ganglion was then perfused with high--Mg$^{2+}$, low--Ca$^{2+}$ saline, containing (in mM) 330 NaCl, 10 KCl, 90 MgCl$_2$, 20 MgSO$_4$, 2 CaCl$_2$, and 10 HEPES (pH 7.6) to prevent synaptic transmission. Intracellular recordings were obtained using Clampex 8.2 software with an Axoclamp 2B amplifier (Axon Instruments, Foster City, CA, USA) in bridge mode using a microelectrode (10-–20 M$\Omega$ resistance) filled with 3 M potassium acetate. The membrane potential was amplified and digitized with a Digidata 1322A board (Axon Instruments) at a rate of 10 kHz.  Current pulses (-0.4 to -1.5 nA) of fixed length (2 to 8 seconds) were triggered externally using a square wave produced with an AFG3021 function generator (Tektronix, Beaverton, OR, USA) which itself was triggered by hand.

We performed experiments on eleven bursting cells from eight animals. In order for a cell to be suitable for analysis, its activity needed to be stable and stationary so we could be confident that spike addition was experimentally induced and not the result of some intrinsic variability. Therefore, we required at least seven bursts to contain the same number of spikes before a stimulation was supplied and, if a mode switch occurred, that it was maintained for seven bursts subsequent to the perturbation. Eight of the eleven cells tested showed activity stationary enough to be analysed, and four of these eight cells showed instances of a sustained mode switch due to perturbation. In the four cells that showed no evidence of multirhythmicity, spike addition could only be maintained for the burst directly following the perturbation and the remaining bursts contained the nominal (pre-stimulus) number of spikes in the active phase. In cells that did show evidence of multirhythmicity, current pulses were not consistently able to induce a switch in bursting mode but our criteria for an attractor switch was met at least one time for each of the four cells, with three out the four cells showing multiple instances of a mode change. We saw no obvious qualitative correlation between characteristics of bursting (periodicity, duty cycle, number of spikes in the active phase, etc) and whether a cell was capable of a mode switch in response to perturbation due to the small sample size and simplicity of our study.

\end{document}